\documentclass[aps,pre,twocolumn,groupedaddress,superscriptaddress,showpacs]{revtex4-1}
\usepackage{amsmath}
\usepackage{graphicx}
\usepackage{float}
\usepackage{xcolor}

\begin{document}

  \title {Nonequilibrium adsorption of 2AnB patchy colloids on substrates}

  \author{C. S. Dias}
   \email{csdias@cii.fc.ul.pt}
    \affiliation{Departamento de F\'{\i}sica, Faculdade de Ci\^{e}ncias da Universidade de Lisboa, P-1749-016 Lisboa, Portugal and Centro de F\'isica Te\'orica e Computacional, Universidade de Lisboa, Avenida Professor Gama Pinto 2, P-1649-003 Lisboa, Portugal}

  \author{N. A. M. Ara\'ujo}
   \email{nuno@ethz.ch}
   \affiliation{Computational Physics for Engineering Materials, IfB, ETH Zurich, Wolfgang-Pauli-Strasse 27, CH-8093 Zurich, Switzerland}

  \author{M. M. Telo da Gama}
   \email{margarid@cii.fc.ul.pt}
    \affiliation{Departamento de F\'{\i}sica, Faculdade de Ci\^{e}ncias da Universidade de Lisboa, P-1749-016 Lisboa, Portugal and Centro de F\'isica Te\'orica e Computacional, Universidade de Lisboa, Avenida Professor Gama Pinto 2, P-1649-003 Lisboa, Portugal}

  \begin{abstract}
We study the irreversible adsorption of spherical $2AnB$ patchy colloids (with two $A$-patches on the poles and $n$ $B$-patches along the equator)
 on a substrate. In particular, we consider dissimilar $AA$, $AB$, and $BB$ binding probabilities. We characterize the patch-colloid network and its 
dependence on $n$ and on the binding probabilities. Two growth regimes are identified with different density profiles and we calculate a growth mode
 diagram as a function of the colloid parameters. We also find that, close to the substrate, the density of the network, which depends on the colloid 
parameters, is characterized by a depletion zone.
  \end{abstract}

  \maketitle

\section{Introduction}

Patchy colloids have functionalized surfaces yielding directional colloid-colloid interactions.\cite{Blaaderen2006,Bianchi2011,Pawar2010,Kretzschmar2011,Grzelczak2010} This feature opens the possibility of controlled bottom-up self-organization of novel materials \cite{Ruzicka2011,Glotzer2010,Chen2011,Romano2011a,Kraft2012} and it has motivated the development of several experimental techniques to produce patchy colloids.\cite{Sacanna2011,Wilner2012,Hu2012,Duguet2011,Pawar2008,Shum2010,Wang2012,He2012} Theoretical studies have been focused on the thermodynamic properties with several numerical techniques being adapted to tackle the patch-colloid cooperative behavior in solution.\cite{Russo2009,Sciortino2011,Sciortino2007,Bianchi2011} 
A range of models for the colloids have been extensively studied and used to represent many complex systems, such as, amphiphilic molecules, colloidal clays, proteins, and DNA nano-assemblies.\cite{Glotzer2004, Doye2007, Sciortino2010, Pawar2010, Glotzer2010, Bianchi2011, Sciortino2011, Rosenthal2011, Ruzicka2011,Romano2012} These studies have revealed that the control over the valence leads to rich phase diagrams with new self-organized phases and interesting properties, such as, glass transitions, colloidal gels, and bigels.\cite{Zaccarelli2005,Zaccarelli2006,Bianchi2006,Russo2010,DelasHeras2012,Varrato2012}

Studies of equilibrium properties provide insight on the possible thermodynamic structures and their stability. However, experimentally, equilibrium structures are very often not accessible. In particular, in the limit of irreversible binding (considered here), the network of colloids is determined by the time sequence of patch-patch bond formation.\cite{Dias2013,Corezzi2012} In general, these kinetically trapped structures are different from the thermodynamic ones. The interest on the identification and characterization of these structures is twofold: designing experimental strategies to avoid them when the equilibrium structures are the goal and the search for novel material properties.

The presence of a substrate improves the degree of controllability over self-organization of patchy particles even under equilibrium conditions.\cite{Gnan2012,Bernardino2012} Recently, the study of the nonequilibrium adsorption on a substrate of two- and three-patch colloids revealed the existence of an optimal fraction of two-patch colloids where the density of the film is maximized.\cite{Dias2013} This result is in contrast to the equilibrium structures where, a monotonic decrease of the density is observed.\cite{Bianchi2006} Several questions about nonequilibrium patchy-colloidal systems are still open, namely, the effect of confinement \cite{Kretzschmar2011}, the inclusion of patterns on the substrate,\cite{Cadilhe2007, Araujo2008} or the possibility of distinct interaction between patches.\cite{Russo2011a} In this work, this last question will be investigated in detail.

A particularly interesting model considers patchy colloids with distinct patch-patch interactions, namely $2AnB$ (two patches of type $A$ and $n$ patches of type $B$), also known as Lisbon colloids. Even the simplest version of this model ($n=1$ with all patches evenly distributed) yields interesting properties in the bulk, such as vapour-liquid transitions and a vanishing critical point. The liquid-vapour critical point was studied as a function of the $AA$, $AB$, and $BB$ interaction energies,\cite{Tavares2009,Tavares2009a} and a rich number of network structures was found to be responsible for this unexpected phase diagram. Both numerical and theoretical results confirm the unusual thermodynamic and percolation properties of this model.\cite{Tavares2010a,Tavares2010} 
More complex $2AnB$ colloids with two patches of type $A$ located on the poles and nine patches of type $B$ located equidistantly on the equator have been studied.\cite{Russo2011,Russo2011a} An equilibrium liquid-vapour binodal was found to be re-entrant with the liquid (a network fluid with a large number of branches) and the vapour (a fluid of linear chains with little or no branching) densities vanishing as the temperature approaches zero. This is in line with the phase diagram predicted by Tlusty and Safran for dipolar fluids.\cite{Tlusty2000} This re-entrant behavior was also found on a lattice model in two and three dimensions.\cite{Almarza2011, Almarza2012}

Until now, the study of $2AnB$ colloids has been restricted to equilibrium properties. This gives an idea of the numerous possibilities that can be expected in the richer nonequilibrium systems. Following this lead, in this work, we will study the nonequilibrium adsorption on an attractive substrate, using our model, proposed in Ref.~\cite{Dias2013}, where the bonding between colloids is irreversible. 
Under this constraint, the system never reaches an equilibrium state. As in previous models of $2AnB$ colloids, \cite{Russo2011,Russo2011a} we consider two patches $A$ in the poles and $n$ patches $B$ around the equator. An example for possible configurations is shown in Fig.~\ref{fig.colloids}(a) with two schematic representations of $2AnB$ colloids with $n=4$ and $n=9$, respectively. We study the effect of varying both the number of patches of type $B$ and the interaction between patches. We find a promising control over the film structure by tunning the $2AnB$ colloids properties. A detailed analysis of the film properties is presented, based on kinetic Monte Carlo simulations. We observe a wide range of coverages of the first layer by changing both the number and the interaction between patches, and the emergence of a density depletion zone near the substrate. A new growth mode diagram depending on the number of $B$-patches and sticking coefficients is proposed.

This article is organized as follows: In Sec.~\ref{sec:model} we introduce the model and the underlying physical considerations, as well as the simulation parameters; In Sec.~\ref{sec:surface} the results for the region close to the substrate are presented and the density depletion in this region is discussed; In Sec.~\ref{sec:liquid} the film growth modes are discussed; Finally, in Sec.~\ref{sec:conclusions} we draw some conclusions.

\section{Model}\label{sec:model}

\begin{figure}[t!]
\centering
    \includegraphics[width=\columnwidth]{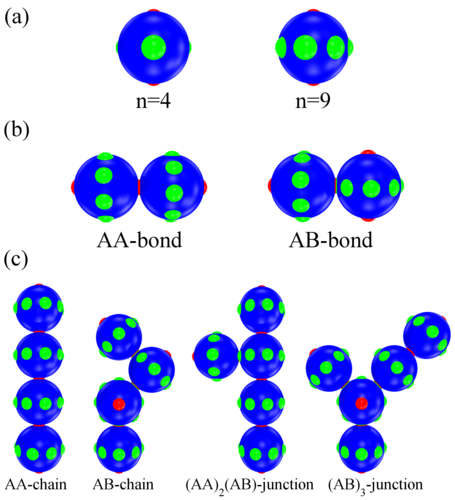} \\
\caption{Schematic representation of $2AnB$ colloids, where $A$ patches are in red and $B$ patches in green. (a) With $n=4$ (left) and $n=9$ (right). (b) Only $AA$ (left) or $AB$ (right) bonds can be established, i.e., $BB$ bonds are neglected. (c) Representative structures formed by assembly, with chains of both $AA$ and $AB$ bonds (left most structures), and junctions of bonds, i.e., multiple bonding of the same colloid (more than two bonds). Junctions are of type $(AA)_x(AB)_y$, where $x=0,...,N_A$ and $y=1,...,N_B+N_A$ ($x+y\geq3$).}
  \label{fig.colloids}
\end{figure}

We consider the stochastic model for the adsorption of patchy particles on a substrate, recently introduced in Ref.~\cite{Dias2013}. Patchy colloids are typically described as spherical particles with patches distributed on their surface, with both a short-range repulsive interaction and a patch-patch attractive interaction. We assume chemical bonding between patches (highly directional and irreversible within the timescale of interest) and we describe the patch-patch binding as a stochastic process, where a bond is established in two different situations: when the patch of the adsorbing colloid is aligned with the patch of a previously adsorbed colloid or when the patch-patch interaction promotes the alignment of the landing colloid with the available binding site. 
An interaction range around the patches is defined such that two colloids effectively interact (and eventually establish a bond) when their interaction ranges overlap. Here, we fix the interaction range within an angle lower than $\pi/6$ rad relative to the center of the patch.\cite{Villar2009} This sets a maximum interaction $\pi\sigma/6$ between patches on the surface of the colloids, where $\sigma$ is the diameter of the colloid, which we consider unity without loss of generality.

To describe the colloids motion in solution, a Brownian dynamics is considered where the collision with the solvent is taken as a Poisson process, such that the time between collisions, $\Delta t$, follows an exponential distribution,
  \begin{equation}
   p(\Delta t)=\exp(-R\Delta t),
   \label{eq.time_dist}
  \end{equation}
where $R$ is the collision rate. At each colloid/solvent collision a new velocity is assigned with the direction uniformly distributed and with absolute value $v$ following the Maxwell-Boltzmann distribution at the thermostat temperature, given by,
  \begin{equation}
   p(v)=\sqrt{\frac{1}{2\pi T}}\exp\left(-\frac{v^2}{2T}\right),
   \label{eq.maxwell_dist}
  \end{equation}
with $T$ the thermostat temperature in units of $k_B/m$, where $m$ is the mass of the colloid and $k_B$ the Boltzmann constant. With this model, it is possible to fine tune the diffusion coefficient $D$, with the proper combination of $T$ and $R$. For the sake of simplicity, we consider a constant diffusion coefficient of $0.1\sigma^2/s$.
 
In the presence of a substrate, there are two characteristic time scales: the inter-arrival time of colloids at the substrate (typically related with the flux); and the binding time, defined as the time necessary for the chemical bond to be established. At low concentration of colloids, the inter-arrival time is considered much larger than the binding time. Therefore, we can assume that colloids arrive one at a time to the substrate and adhere instantaneously. For the purpose of this study, the interaction of a colloid with the substrate is isotropic and they adsorb with no defined direction. 

The process of establishing a chemical bond is a thermally activated process, characterized by an activation barrier. In general, this barrier is different for $AA$, $AB$, and $BB$ bonds. Here, following previous studies of $2AnB$ colloids,\cite{Russo2011a,Russo2011} we consider that the activation barrier for $BB$ is so much higher than the others, that such bonds can, in practice, be neglected. We define the binding probability for $AA$ and $AB$ bonds as $P_{AA}$ and $P_{AB}$. For simplicity, we consider the limit $P_{AA}=1$ and define the sticking coefficient $r_{AB}=P_{AB}/P_{AA}$. The two possible bonds are represented in Fig.~\ref{fig.colloids}(b).

The parameters $n$ and $r_{AB}$ are related. The larger their values, the higher the probability that an $AB$ bond is formed. Yet, as we show below, a different quantitative and qualitative dependence on each parameter is observed. Figure~\ref{fig.colloids}(c) shows some possible local arrangements obtained with $2AnB$ colloids. The examples are: a sequence of consecutive $AA$ bonds, always resulting in a linear chain of colloids; $AB$-chains, formed by sequences of $AB$-bonds, which are rarely linear (linear $AB$-chains are only geometrically possible for even values of $n$) whose shape depends strongly on $n$; and $(AA)_x(AB)_y$-junctions, where $x=0,...,N_A$ and $y=1,...,N_B+N_A$ ($x+y\geq3$), which are colloids with more than two bonds. 

In a previous work for colloids with identical patches,\cite{Dias2013} we found three distinct regimes in the density profile: the surface layer, where the effect of the substrate, on the colloid network, plays a major role; the liquid film, where the density of the film reaches a constant value during growth; and the interfacial region, which is mainly related to the roughness of the film. In this article, we focus on the surface layer and liquid film for $2AnB$ colloids.

\section{Surface Layer}\label{sec:surface}

The layer of colloids directly in contact with the substrate is the precursor film. As we show here, for a patchy-colloid network with strongly anisotropic colloid-colloid interaction, the organization of the first layer affects the growth of the remaining structure. For nonequilibrium adsorption of patchy colloids with identical patches, it has been shown\cite{Dias2013} that the density of the first layer remains constant, independently of the system size, but increases with the diffusion coefficient $D$ of the colloid.

\begin{figure}[t!]
\centering
    \includegraphics[width=\columnwidth]{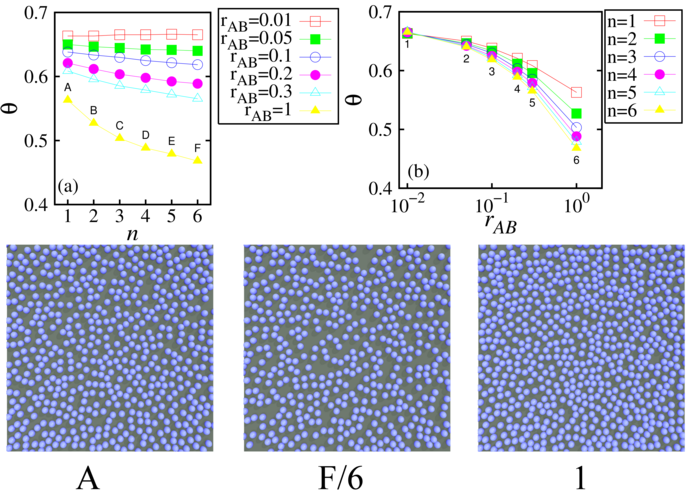} \\
    \caption{(a) Coverage of the substrate, $\theta$ (number of particles directly adsorbed on the substrate per unit area), as a function of the number of $B$-patches, $n$, for values of $r_{AB}$ between $0.01$ and $1$. (b) Coverage of the substrate, $\theta$, as a function of the sticking coefficient, $r_{AB}$, for values of $n=\{1,2,3,4,5,6\}$. (bottom) Snapshots of the first adsorbed layer (directly in contact with the substrate) for three different points on the plots (a) and (b), namely, 1, A, and F/6, where the dependence of the density on $n$ and $r_{AB}$ is observed. Simulations performed on a substrate of lateral size 32 in units of particle diameter.}
    \label{fig.firstlayer}
\end{figure}

The coverage of the first layer, i.e., the number of particles per unit of area, as a function of $n$ and $r_{AB}$ is shown in Figs.~\ref{fig.firstlayer}(a)~and~(b), respectively. It is observed, in Fig.~\ref{fig.firstlayer}(a), that for $r_{AB}\approx 1$, the coverage of the first layer decreases with $n$, while for $r_{AB}\ll1$, it remains constant. At low $r_{AB}$ the probability of binding to a previously adsorbed particle is very low and, consequently, particles are able to diffuse towards the substrate and bind directly to it. 
This binding is solely limited by the colloid-colloid excluded volume interaction and, in the limit of vanishing $r_{AB}$, the coverage of the substrate by particles directly in contact is expected to resemble the Random Sequential Adsorption model,\cite{Evans1993a} with a jamming coverage, $\theta_\infty=0.6969$. The effect of the number of patches will then be negligible for $r_{AB}\ll1$ (like, e.g., $r_{AB}=0.01$ in Fig.~\ref{fig.firstlayer}(a)). 

In general, for any value of $r_{AB}$, a decrease of the coverage of the first layer with $n$ is observed. This is related with the increase in the probability that a particle binds to the network, thus enhancing the screening effect. A similar qualitative behavior is observed with increasing $r_{AB}$. This dependence on $n$ and $r_{AB}$ is also visible in the snapshots of the first layer, Fig.~\ref{fig.firstlayer}, for limiting values of $n$ and $r_{AB}$, where the density at low $n$ and $r_{AB}$ is higher than at higher values.

\begin{figure}[t!]
\centering
    \includegraphics[width=\columnwidth]{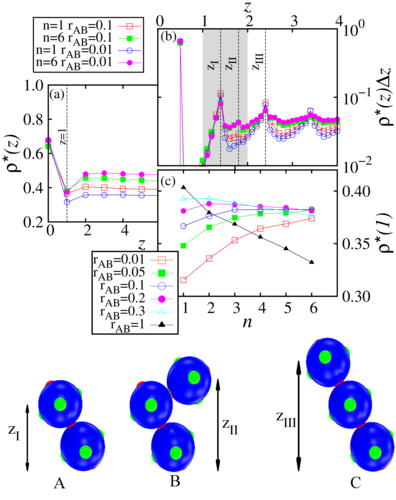} \\
    \caption{(a) Density profile of the colloidal network, close to the substrate, from $z=0$ to $z=10$ and various values of $n$ and $r_{AB}$, with a binning of $\Delta z=1$. From bottom to top  $\{N_B=1,r_{AB}=0.01\}, \{N_B=1,r_{AB}=0.1\}, \{N_B=6,r_{AB}=0.1\},$ and $\{N_B=6,r_{AB}=0.01\}$. (b) Density profile of the colloidal network on a substrate, with binning of $\Delta z=0.1$ from  $z=0$ to $z=4$, for the same parameters and symbols of plot (a). The shadowed region corresponds to the volume used for the calculation of the density at $1<z<2$ in plot (a). Labels A, B, and C indicate the first three peaks in the density above the layer directly in contact with the substrate. (c) Density of the region at $1<z<2$ as a function of $r_{AB}$ for $n=1$ (open squares), $2$ (solid squares), $3$ (open circles), $4$ (solid circles), $5$ (open triangles), and $6$ (solid triangles). Simulations performed on a substrate of lateral size $L=32$, in units of particle diameter, and a total of $61440$ deposited colloids. 
(bottom) Schematic representation of the structures at the peaks marked in (b).}
    \label{fig.layers123}
\end{figure}

For patchy colloids with identical patches, a decrease of the density with the distance to the substrate is observed for the surface layer.\cite{Dias2013} However, this is not the case for $2AnB$ colloids, with distinct patches. Figure~\ref{fig.layers123}(a) shows several density profiles obtained with these particles, where a minimum is observed for $z<2$. The presence and position of this minimum is robust with respect to the values of $n$ and $r_{AB}$. Below, we discuss this behavior in detail.

Figure~\ref{fig.layers123}(a) shows the density profile for some values of $r_{AB}$ and $n$ where the minimum is observed. An explanation may be drawn from Fig.~\ref{fig.layers123}(b), where a density profile is shown, with a binning of $\Delta z=0.1$. Here, a narrow peak appears at a distance of $1.3<z<1.4$ from the substrate ($z_I$ in the figure). This is a direct consequence of the high density of the first layer and the low probability of adsorption on $B$-patches. 
The scheme A in the bottom of Fig.~\ref{fig.layers123} illustrates the most common configuration of the colloid connected to the one that is directly bound to the substrate. For an isotropic orientation of colloids in the first layer, the probability of finding a colloid near the normal to the substrate is very low and, due to excluded volume, adsorption of colloids on patches that form an angle lower than $\pi/2$ rad with the substrate is also limited. Considering a coverage around $0.65$ particles per unit of area, we obtain an area of approximately $1.54\sigma^2$ per particle. The average distance between particles can be estimated from the diameter of such regions, yielding $\Delta r\approx 1.4\sigma$. The minimum height of a colloid, not directly in contact with the substrate, would correspond to a particle resting on top of two other particles at this distance. In that case, a minimum $z\approx1.21$ is obtained. Since bonds can only be established along the direction of the patches, the maximum is 
observed in the range $1.3<z<1.4$.

What enhances this behavior with $2AnB$ colloids is the lower probability to bind to $B$-patches, since $BB$ bonds are absent and $r_{AB}\leq1$. However, when an attempt of adsorption on a $B$-patch is successful, the new bond works as the seed to a lateral chain, which consequently screens the adsorption on the region $1<z<2$ (scheme B in the bottom of Fig.~\ref{fig.layers123}). This few successful adsorptions on lateral $B$ patches are represented in Fig.~\ref{fig.layers123}(b) as a small peak just bellow $z=2$ ($z_{II}$). When $r_{AB}\approx1$ the colloidal network resembles the one for colloids with identical patches, where the density decreases monotonically with the distance to the substrate due to branching. In this limit, the effect of the dense coverage of the first layer is negligible, as new colloids can adsorb equally on any patch. 

\begin{figure}[t!]
\centering
    \includegraphics[width=\columnwidth]{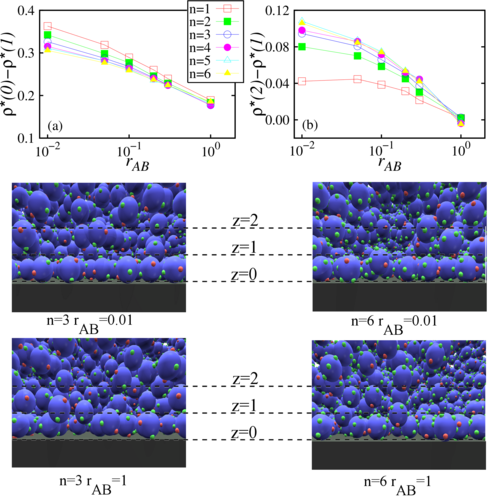} \\
    \caption{(a) Density difference, $\rho^*(0)-\rho^*(1)$, corresponding to $0<z<1$ and $1<z<2$, respectively, as a function of $r_{AB}$ for values of $n=\{1,2,3,4,5,6\}$. (b) Density difference, $\rho^*(2)-\rho^*(1)$, corresponding to $2<z<3$ and $1<z<2$, respectively, as a function of $r_{AB}$ for values of $n=\{1,2,3,4,5,6\}$. (bottom) Snapshots for, from left to right and top to bottom, $\{N_B=3,r_{AB}=0.01\}, \{N_B=6,r_{AB}=0.01\}, \{N_B=3,r_{AB}=1\},$ and $\{N_B=6,r_{AB}=1\}$. Results averaged over 100 samples on a square substrate of lateral size $L=32$ in units of the particle diameter and a total of $61440$ deposited colloids.}
    \label{fig.layers_gap}
\end{figure}

Figure~\ref{fig.layers123}(c) shows the density of the shadowed region in Fig.~\ref{fig.layers123}(b) as a function of $n$, for different values of $r_{AB}$. For lower $r_{AB}$, the density increases with $n$. By contrast, for higher $r_{AB}$, it decreases with $n$. For low values of $r_{AB}$ the increase in the density is related to larger branching and lateral growth of the film, since the first layer has a constant density regardless of $n$, as stated above. The higher the value of $n$ the higher the branching rate and, consequently, the number of lateral adsorptions on colloids of the second layer. The number of possible bonds along the equator is limited by the excluded volume interaction among colloids. Thus, it is expected that above $n=4$, the efficiency of adsorption does not improve with $n$. For high values of $r_{AB}$, the density of the shadowed region in Fig.~\ref{fig.layers123}(b) varies with the density of the first layer.

The control of the density in the depletion zone can be achieved through the choice of $n$ and $r_{AB}$ (see Fig.~\ref{fig.layers_gap}). The density close to the substrate ($z<1$) is always larger than in the second layer ($1<z<2$). Due to the higher probability to stick to the substrate than to any patch (see Fig.~\ref{fig.layers_gap}(a)). The second layer ($1<z<2$) can have a density higher or lower than the third ($2<z<3$) depending on $n$ and $r_{AB}$. This is a direct consequence of the distinct patch-patch interaction since, as observed in Fig.~\ref{fig.layers_gap}(b), for higher values of $r_{AB}$, the depletion disappears, and the behavior is similar to that observed for colloids with identical patches.\cite{Dias2013} 
Snapshots of the film close to the substrate show a reduced density for $1<z<2$, more evident for low values of $r_{AB}$ (see top snapshots in Fig.~\ref{fig.layers_gap}), and the effect of lateral chains (scheme B in Fig.~\ref{fig.layers123}).

\section{Liquid film}\label{sec:liquid}

\begin{figure}[t!]
\centering
    \includegraphics[width=\columnwidth]{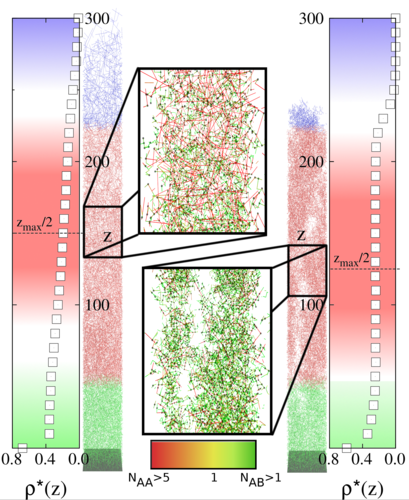} \\
    \caption{Density profile of the colloidal network on a substrate showing three different regimes: surface layer (green, bottom), liquid film (red, middle), and interfacial region (blue, top). Left: Reduced density, $\rho^*(z)$ (number of particles per unit of volume), as a function of $z$, for $n=1$ and $r_{AB}=0.01$; Right: Density, $\rho^*(z)$ as a function of $z$, for $n=1$ and $r_{AB}=1$; (center) Magnified region of the colloidal network (top) $n=1$ and $r_{AB}=0.01$, and (bottom) $n=1$ and $r_{AB}=1$. Color gradient indicates $AA$-chains (red) and $AB$-chains (green), with junctions represented as black spheres. Both cases after the adsorption of $40960$ particles, averaged over 100 independent realizations. $z_{max}/2$ is the value of half the maximum thickness of the film.}
    \label{fig.density_regimes}
\end{figure}

For the adsorption of colloids with identical patches, it was observed that the lateral growth of the colloidal network leads to a saturation in the film density ($\rho=\rho_l$), due to the finite size of the substrate (see right snapshot in Fig.~\ref{fig.density_regimes}). Surprisingly, with distinct patches, we found for low values of $r_{AB}$ a new type of growth where no saturation of the density is observed (left snapshot in Fig.~\ref{fig.density_regimes}).
Figure~\ref{fig.density_regimes} shows snapshots and density profiles for the two different growth modes, where the difference in the film structure is visible. For higher values of $r_{AB}$ we observe a higher branching rate due to the adsorption on both $A$ and $B$ patches (bottom zoom of Fig.~\ref{fig.density_regimes}). The junctions consist mainly of $AB$-bonds, which can initiate the formation of $AB$-chains (Fig.~\ref{fig.colloids}(c)). These chains might have different geometries as pictured in Fig.~\ref{fig.colloids}(c). By contrast, for lower values of $r_{AB}$, the ratio and size of $AA$-chains (Fig.~\ref{fig.colloids}(c)) increases with the number of deposited particles (top zoom of Fig.~\ref{fig.density_regimes}), which is associated with a decrease in the density, since a compaction of the film is hindered by the linear $AA$-chains.

The different behavior at $r_{AB}\ll1$ is a consequence of an increase in the relative density of $AA$ over $AB$ bonds. By contrast with $2AnB$ colloids in equilibrium, where bonds are reversible and a low value of $r_{AB}$ leads to a negligible number of $AB$-bonds, under nonequilibrium conditions, even for low $r_{AB}$, $AB$-bonds still contribute to the structure of the film. In equilibrium, for low values of $r_{AB}$, the structure consists mostly of $AA$-chains\cite{Russo2011a} at an almost constant density around the equilibrium value. Out of equilibrium, for a network of $2AnB$ colloids in the limit of $r_{AB}\ll1$, the density of bonds of type $AB$ decreases with the number of adsorbed particles. This effect is due to the increasing dominance of $AA$ chains, which will screen available patches inside the film and, consequently, keep the density of the film low. The large number of $B$ patches available for an $AB$-bond are shadowed by particles above, and remain unreachable for newly incoming 
colloids.

\begin{figure}[t!]
\centering
    \includegraphics[width=\columnwidth]{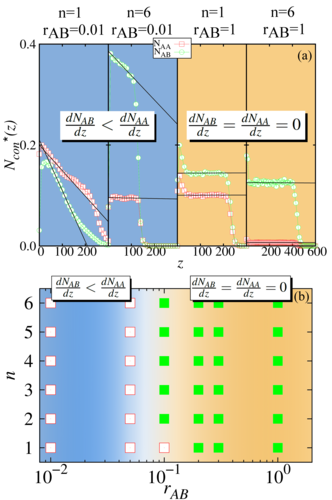} \\
    \caption{Analysis of the growth modes for different values of $n$ and $r_{AB}$. (a) Number of $AA$- (red open squares) and $AB$-bonds (green solid squares) per unit volume, $N_{con}*(z)$, as a function of $z$, for values of $r_{AB}$ and $n$ from left to right: $\{r_{AB}=0.01; N_B=1\}$, $\{r_{AB}=0.01; N_B=6\}$, $\{r_{AB}=1; N_B=1\}$, and $\{r_{AB}=1; N_B=6\}$. Solid lines indicate the derivative of the function in the liquid film regime. (b) Growth mode diagram for values of $n$ as a function of $r_{AB}$, for the Chain Dominated Regime (red open squares, left blue region) and the Junction Dominated Regime (green solid squares, right orange region). Results for $61440$ deposited particles on a substrate with $L=32$ averaged over 100 samples.}
    \label{fig.growth_diagram}
\end{figure}

For colloids with identical patches, the liquid regime was defined as the regime, above the surface layer, where the density saturates.\cite{Dias2013} Here, we found a growth mode where the density of the film does not saturate. Despite a continuous decrease of the density (left of Fig.~\ref{fig.density_regimes}), a liquid regime can be identified as the region where the finite size of the substrate plays a relevant role. This decrease of the film density with the number of deposited particles is a result of the dominance of the rate of formation of $AA$-chains over the rate of formation of junctions or $AB$-bonds. 
This behavior is evident from the derivative of the density of $AA$- and $AB$-bonds, $\frac{dN_{AA}}{dz}$ and $\frac{dN_{AB}}{dz}$, inside the liquid film regime, represented by the solid lines in Fig.~\ref{fig.growth_diagram}(a), where $\frac{dN_{AA}}{dz}>\frac{dN_{AB}}{dz}$. We name this growth mode Chain Dominated Growth (CDG). This is in sharp contrast with the second growth mode, where the density saturates (right of Fig.~\ref{fig.density_regimes}) and $\frac{dN_{AA}}{dz}=\frac{dN_{AB}}{dz}=0$. We name this mode Junction Dominated Growth (JDG). To compute the derivative of the density of $AA$- and $AB$-bonds, the liquid regime region needs to be defined precisely. For the JDG mode, the liquid regime is the region where the density of bonds remains constant. In the CDG mode, as a result of the decreasing density, the density of bonds exhibits a maximum (see Fig.~\ref{fig.growth_diagram}(a)). 
As we go away from the substrate, the density of bonds initially increases due to the absence of binding to the substrate and the low density of the depletion zone close to the substrate. The maximum can then be considered a transition point between the surface layer and the liquid regime, and can be used as a reference for the measurements in the CDG mode.

To characterize the growth-mode diagram and its dependence on $n$ and $r_{AB}$ we compute  the derivative of the density of $AA$ and $AB$ bonds as a function of $z$. In Fig.~\ref{fig.growth_diagram}(a) one can see that for the CDG mode the density of $AB$ bonds decreases with $z$, and for lower values of $n$ the density of $AA$ bonds also decreases. The overall decay of both can be related to the decrease of the film density. In the CDG, however, the rate of decrease of $AB$-bonds is higher than that of $AA$-bonds. We can then characterize the growth modes as,
\begin{eqnarray}
 &\frac{dN_{AB}}{dz}<\frac{dN_{AA}}{dz}& \text{ CDG} \nonumber \\
\text{and} \\
 &\frac{dN_{AB}}{dz}=\frac{dN_{AA}}{dz}=0& \text{ JDG.} \nonumber
 \label{eq.growth_regimes}
\end{eqnarray}

A growth diagram is shown in Fig.~\ref{fig.growth_diagram}(b) for the range of $n$ and $r_{AB}$ investigated. For $r_{AB}<0.1$ the growth is CDG and for $r_{AB}\geq0.1$ it is JDG. For $n=1$ and $r_{AB}=0.1$ the growth mode is still CDG.

\begin{figure}[t!]
\centering
    \includegraphics[width=\columnwidth]{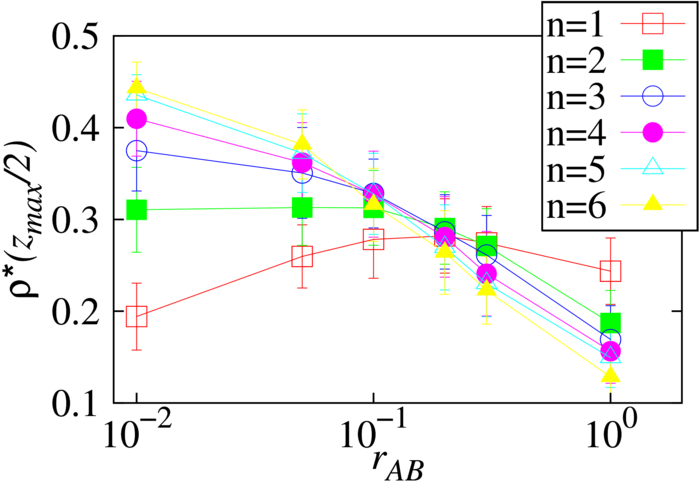} \\
    \caption{Density at $z=z_{max}/2$ as a function of $r_{AB}$ for different values of $n$: one (open squares), two (solid squares), three (open circles), four (solid circles), five (open triangles), and six (solid triangles). Results for $61440$ deposited particles on a substrate with $L=32$ averaged over 100 samples.}
    \label{fig.half_density}
\end{figure}

For $n<4$, $r_{AB}$ has a large effect on the formation of junctions, however this effect disappears for $n$ larger than four due to steric limitations. In the limiting case of $r_{AB}\ll1$, there is a strong dependence of the overall density on $n$, so, for lower values of $n$, the number of unsuccessful binding attempts is higher and the density increases through the diffusion of colloids into the fjords of the film. For the case of $r_{AB}\approx1$, a higher number of successful bonds leads to a larger ramification of the film. 
In the limit of $r_{AB}\approx1$, $n$ will have an effect similar to $r_{AB}$, i.e., it increases the rate of bonding, as most of the patches are not available for binding due to the screening by newly deposited colloids. Since for CDG no saturation of the density is observed, the density at half the thickness of the film, $\rho(z_{max}/2)$ was measured (see Fig.~\ref{fig.half_density}). Taking $z_{max}/2$ as defined in Fig.~\ref{fig.density_regimes}, for $n>2$, the density decreases with $r_{AB}$. 
This is a consequence of a higher ratio of $AB$-bonds, which favors the formation of junctions and $AB$-chains, and, consequently, promotes ramification. Thus, it results in the appearance of regions of very low density. For $n\leq2$, the density exhibits a maximum at the transition between the CDG and JDG modes. This is due to the effect of $AA$-chains, which, for the CDG mode, dominates at lower $n$. We note that, in the CDG mode, the density of the film is expected to vanish in the limit $z\rightarrow\infty$.

\begin{figure}[t!]
\centering
     \includegraphics[width=\columnwidth]{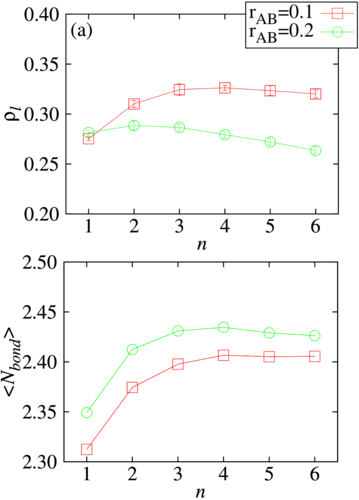} \\
    \caption{Analysis of the liquid film density near the transition between growth regimes. (a) Liquid film density, $\rho_l$, as a function of $n$ for $r_{AB}=0.1$ (open squares) and $r_{AB}=0.2$ (open circles). (b) Average number of bonds per colloid (colloids with more than one bond), $<N_{bond}>$, as a function of $n$ for $r_{AB}=0.1$ (open squares) and $r_{AB}=0.2$ (open circles). Results for $61440$ deposited particles on a substrate with $L=32$ averaged over 100 samples.}
    \label{fig.plateau_branch}
\end{figure}

The results of Fig.~\ref{fig.half_density}, at values of $r_{AB}$, inside the JDG mode, reveal a non-uniform dependence of the density on $n$. This unusual behavior is observed at values of $r_{AB}=0.1$ and $r_{AB}=0.2$. Due to large fluctuations of the density at $z_{max}/2$ (from the error bars of Fig.~\ref{fig.half_density}), the saturation density $\rho_l$ was chosen as the variable to characterize the film. Figure~\ref{fig.plateau_branch}(a) shows that the film density exhibits a maximum that depends on $r_{AB}$ and $n$. For $r_{AB}=0.1$ the maximum is found at $n=4$ and for $r_{AB}=0.2$ at $n=2$. For these pairs of $r_{AB}$ and $n$, the film grows according to the CDG mode. In this growth mode, favored by junctions, the maximum in the density is driven by the maximum number of bonds per particle.  
This is the case for $n=4$, as seen in Fig.~\ref{fig.plateau_branch}(b), where the average number of bonds per particle, with more than one bond, is plotted. We observe that for both $r_{AB}=0.1$ and $r_{AB}=0.2$ the maximum in the average number of bonds occurs for $n\approx4$, implying that the configuration with two patches in the poles and four around the equator (octahedral geometry) maximizes the formation of junctions. As a conclusion, an increase in $n$ does not necessarily promote junctions, which explains the maximum in Fig.~\ref{fig.plateau_branch}(a). For $r_{AB}=0.2$, however, branching and formation of voids shift the density maximum towards lower values of $n$ (around two).

\section{Conclusions}\label{sec:conclusions}

We have studied the structural properties of colloidal networks obtained from the irreversible adsorption of $2AnB$ colloids on a substrate. This is the first study of nonequilibrium growth of patchy colloids with a heterogeneous distribution of patches, namely, two types of patches restricted to the poles and the equator. We have shown that these $2AnB$ colloids are endowed with new tunable parameters for the control of the network properties.

We found that $2AnB$ colloids exhibit different structures from those observed in $(2+n)A$ colloids both at the surface and within the liquid film. A depletion of particles at one particle diameter from the substrate is observed. We also show that the coverage of the substrate can be controlled through the number of patches in the equator and the unlike patch-patch interaction. For simplicity, we have considered an isotropic particle-substrate interaction. In general, some anisotropy is expected due to the presence of the patches. We have shown\cite{Dias2013} that, when only one type of patch is considered, the liquid film regime is not affected by the details of the interaction with the substrate. Nevertheless, differences have been observed in the surface-layer regime. In the presence of two types of patches a richer surface-layer regime is expected depending on the selectivity of the particle-substrate interactions.

We summarized the different growth modes in a two-parameter diagram, namely $n$ and $r_{AB}$. We have identified two growth modes that we named Chain Dominated Growth (CDG) and Junction Dominated Growth (JDG). The structure of the film in the two modes is intrinsically different, for example: in the CDG mode the density of the film decreases with the thickness while for the JDG mode it remains constant. We also propose to control the overall film density by tunning the number of patches and their interaction.

These networks might be of relevance in the fields of microfluidics and filtering. For this purpose, the experimental realization of $2AnB$ colloids is necessary and could be achieved by using triblock colloids\cite{Chen2011} or colloidal particles functionalized by DNA.\cite{Geerts2010,Wang2012} As in previous work,\cite{Dias2013} we have focused on the nonequilibrium properties in the limit of irreversible binding, usually associated with low temperatures, where the obtained colloidal structures do not change over time. However, on very long timescales or at higher temperatures, the possibility of breaking bonds can no longer be neglected and the film is expected to relax towards the equilibrium structure. The time evolution of these structures and the identification of the characteristic timescales of relaxation are interesting open questions which are left for future work.

\acknowledgments{ We acknowledge financial support from the
Portuguese Foundation for Science and Technology (FCT) under Contracts
nos. EXCL/FIS-NAN/0083/2012, PEst-OE/FIS/UI0618/2011, and
PTDC/FIS/098254/2008, and also helpful comments from Francisco de los
Santos. This work was also supported (NA) by grant number FP7-319968 
of the European Research Council}

  \bibliography{colloids_lisbon}

\end{document}